\documentstyle[10pt,epsf,epsfig,dp_delphititle]{dp_delphi}
\makeindex
\def\DpPaperGroup{EP}
\def\DpPaperRef{2000-030}
\def\DpDate{16 February 2000}
\def\DpAuthors{DELPHI Collaboration}
\def\DpSubmit{(Phys. Lett. B478(2000)14)}
\def\DpTitle{{Determination of \boldmath $|V_{ub}|/|V_{cb}|$\\ 
with DELPHI at LEP}}
\def\DpComment{ }
\def\DpEMail{ }
\begin{document}
%%%%%%%%%%%%%%%%%%%%%%%%%% They are a problem with Coll.Sty ?
\makeatletter
%%%CD\input{dp_system:coll.sty}
% Collapse citation numbers to ranges.  Non-numeric and undefined labels
% are handled.  No sorting is done.  E.g., 1,3,2,3,4,5,foo,1,2,3,?,4,5
% gives 1,3,2-5,foo,1-3,?,4,5
\newcount\@tempcntc
\def\@citex[#1]#2{\if@filesw\immediate\write\@auxout{\string\citation{#2}}\fi
  \@tempcnta\z@\@tempcntb\m@ne\def\@citea{}\@cite{\@for\@citeb:=#2\do
    {\@ifundefined
       {b@\@citeb}{\@citeo\@tempcntb\m@ne\@citea\def\@citea{,}{\bf ?}\@warning
       {Citation `\@citeb' on page \thepage \space undefined}}%
    {\setbox\z@\hbox{\global\@tempcntc0\csname b@\@citeb\endcsname\relax}%
     \ifnum\@tempcntc=\z@ \@citeo\@tempcntb\m@ne
       \@citea\def\@citea{,}\hbox{\csname b@\@citeb\endcsname}%
     \else
      \advance\@tempcntb\@ne
      \ifnum\@tempcntb=\@tempcntc
      \else\advance\@tempcntb\m@ne\@citeo
      \@tempcnta\@tempcntc\@tempcntb\@tempcntc\fi\fi}}\@citeo}{#1}}
\def\@citeo{\ifnum\@tempcnta>\@tempcntb\else\@citea\def\@citea{,}%
  \ifnum\@tempcnta=\@tempcntb\the\@tempcnta\else
   {\advance\@tempcnta\@ne\ifnum\@tempcnta=\@tempcntb \else \def\@citea{--}\fi
    \advance\@tempcnta\m@ne\the\@tempcnta\@citea\the\@tempcntb}\fi\fi}
 
\makeatother
%%%%%%%%%%%%%%%%%%%%%%%%%% ??????????????????????????????????
% Generate the title page
\begin{titlepage}
\pagenumbering{roman}
\CERNpreprint{\DpPaperGroup}{\DpPaperRef} % Reference of the paper
\date{{\small\DpDate}} % Date of the paper
\title{\DpTitle} % Title of the paper
\address{\DpAuthors} % General name of the author(s)
\begin{shortabs} % Start the abstract
\noindent
%   abstract.tex
%
\noindent

The ratio of the CKM quark-mixing matrix elements $|V_{ub}|/|V_{cb}|$ 
has been measured using $B$ hadron semileptonic decays. 
The analysis 
%is based on a novel technique, recently proposed by theoreticians, that 
uses the reconstructed mass $M_X$ of the secondary hadronic system 
produced in association with an identified lepton. 
Since $B \rightarrow X_u \ell \bar \nu$ transitions are characterised by 
hadronic masses below those of the $D$ mesons produced in 
$B \rightarrow X_c \ell \bar \nu$ transitions, 
events with a reconstructed value of $M_X$ significantly below the $D$ mass
are selected.
Further signal enrichments are obtained using the 
topology of reconstructed decays and hadron identification. 
A fit to the numbers of decays in the $b\to u$ enriched and depleted samples
with $M_X$ above and below 1.6 GeV/$c^2$ 
and to the shapes of the lepton energy distribution in the $B$ 
rest frame gives 
$|V_{ub}|/|V_{cb}|=
\mathrm{0.103^{+0.011}_{-0.012}~(stat.)\pm0.016~(syst.)\pm0.010~(model)}$
%The model dependance of $b \rightarrow u \ell \bar \nu$ 
%is significantly reduced with this method. 
and, correspondingly, a charmless semileptonic $B$ decay branching fraction 
of BR($B \rightarrow X_u \ell \bar \nu) = \mathrm{(1.57 \pm 0.35~(stat.)
\pm0.48~(syst.)\pm0.27~(model))\times 10^{-3}}$.
\end{shortabs}
\vfill
\begin{center}
\DpSubmit \ \\ % Horrible hack to allow to have DpSubmit empty
\DpComment \ \\
\DpEMail \ \\
\end{center}
\vfill
\clearpage
\headsep 10.0pt
\addtolength{\textheight}{10mm}
\addtolength{\footskip}{-5mm}
\begingroup
% Commands to process the author names
%
\newcommand{\DpName}[2]{\hbox{#1$^{\ref{#2}}$},\hfill}
\newcommand{\DpNameTwo}[3]{\hbox{#1$^{\ref{#2},\ref{#3}}$},\hfill}
\newcommand{\DpNameThree}[4]{\hbox{#1$^{\ref{#2},\ref{#3},\ref{#4}}$},\hfill}
\newskip\Bigfill \Bigfill = 0pt plus 1000fill
\newcommand{\DpNameLast}[2]{\hbox{#1$^{\ref{#2}}$}\hspace{\Bigfill}}
%
%\small
\footnotesize
\noindent
\DpName{P.Abreu}{LIP}
\DpName{W.Adam}{VIENNA}
\DpName{T.Adye}{RAL}
\DpName{P.Adzic}{DEMOKRITOS}
\DpName{Z.Albrecht}{KARLSRUHE}
\DpName{T.Alderweireld}{AIM}
\DpName{G.D.Alekseev}{JINR}
\DpName{R.Alemany}{VALENCIA}
\DpName{T.Allmendinger}{KARLSRUHE}
\DpName{P.P.Allport}{LIVERPOOL}
% ...
\normalsize
\endgroup
\titlefoot{Department of Physics and Astronomy, Iowa State
     University, Ames IA 50011-3160, USA
    \label{AMES}}
\titlefoot{Physics Department, Univ. Instelling Antwerpen,
     Universiteitsplein 1, B-2610 Antwerpen, Belgium \\
     \indent~~and IIHE, ULB-VUB,
     Pleinlaan 2, B-1050 Brussels, Belgium \\
     \indent~~and Facult\'e des Sciences,
     Univ. de l'Etat Mons, Av. Maistriau 19, B-7000 Mons, Belgium
    \label{AIM}}
\titlefoot{Physics Laboratory, University of Athens, Solonos Str.
     104, GR-10680 Athens, Greece
    \label{ATHENS}}
\titlefoot{Department of Physics, University of Bergen,
     All\'egaten 55, NO-5007 Bergen, Norway
    \label{BERGEN}}
\titlefoot{Dipartimento di Fisica, Universit\`a di Bologna and INFN,
     Via Irnerio 46, IT-40126 Bologna, Italy
    \label{BOLOGNA}}
\titlefoot{Centro Brasileiro de Pesquisas F\'{\i}sicas, rua Xavier Sigaud 150,
     BR-22290 Rio de Janeiro, Brazil \\
     \indent~~and Depto. de F\'{\i}sica, Pont. Univ. Cat\'olica,
     C.P. 38071 BR-22453 Rio de Janeiro, Brazil \\
     \indent~~and Inst. de F\'{\i}sica, Univ. Estadual do Rio de Janeiro,
     rua S\~{a}o Francisco Xavier 524, Rio de Janeiro, Brazil
    \label{BRASIL}}
\titlefoot{Comenius University, Faculty of Mathematics and Physics,
     Mlynska Dolina, SK-84215 Bratislava, Slovakia
    \label{BRATISLAVA}}
\titlefoot{Coll\`ege de France, Lab. de Physique Corpusculaire, IN2P3-CNRS,
     FR-75231 Paris Cedex 05, France
    \label{CDF}}
\titlefoot{CERN, CH-1211 Geneva 23, Switzerland
    \label{CERN}}
\titlefoot{Institut de Recherches Subatomiques, IN2P3 - CNRS/ULP - BP20,
     FR-67037 Strasbourg Cedex, France
    \label{CRN}}
\titlefoot{Now at DESY-Zeuthen, Platanenallee 6, D-15735 Zeuthen, Germany
    \label{DESY}}
\titlefoot{Institute of Nuclear Physics, N.C.S.R. Demokritos,
     P.O. Box 60228, GR-15310 Athens, Greece
    \label{DEMOKRITOS}}
\titlefoot{FZU, Inst. of Phys. of the C.A.S. High Energy Physics Division,
     Na Slovance 2, CZ-180 40, Praha 8, Czech Republic
    \label{FZU}}
\titlefoot{Dipartimento di Fisica, Universit\`a di Genova and INFN,
     Via Dodecaneso 33, IT-16146 Genova, Italy
    \label{GENOVA}}
\titlefoot{Institut des Sciences Nucl\'eaires, IN2P3-CNRS, Universit\'e
     de Grenoble 1, FR-38026 Grenoble Cedex, France
    \label{GRENOBLE}}
\titlefoot{Helsinki Institute of Physics, HIP,
     P.O. Box 9, FI-00014 Helsinki, Finland
    \label{HELSINKI}}
\titlefoot{Joint Institute for Nuclear Research, Dubna, Head Post
     Office, P.O. Box 79, RU-101 000 Moscow, Russian Federation
    \label{JINR}}
\titlefoot{Institut f\"ur Experimentelle Kernphysik,
     Universit\"at Karlsruhe, Postfach 6980, DE-76128 Karlsruhe,
     Germany
    \label{KARLSRUHE}}
\titlefoot{Institute of Nuclear Physics and University of Mining and Metalurgy,
     Ul. Kawiory 26a, PL-30055 Krakow, Poland
    \label{KRAKOW}}
\titlefoot{Universit\'e de Paris-Sud, Lab. de l'Acc\'el\'erateur
     Lin\'eaire, IN2P3-CNRS, B\^{a}t. 200, FR-91405 Orsay Cedex, France
    \label{LAL}}
\titlefoot{School of Physics and Chemistry, University of Lancaster,
     Lancaster LA1 4YB, UK
    \label{LANCASTER}}
\titlefoot{LIP, IST, FCUL - Av. Elias Garcia, 14-$1^{o}$,
     PT-1000 Lisboa Codex, Portugal
    \label{LIP}}
\titlefoot{Department of Physics, University of Liverpool, P.O.
     Box 147, Liverpool L69 3BX, UK
    \label{LIVERPOOL}}
\titlefoot{LPNHE, IN2P3-CNRS, Univ.~Paris VI et VII, Tour 33 (RdC),
     4 place Jussieu, FR-75252 Paris Cedex 05, France
    \label{LPNHE}}
\titlefoot{Department of Physics, University of Lund,
     S\"olvegatan 14, SE-223 63 Lund, Sweden
    \label{LUND}}
\titlefoot{Universit\'e Claude Bernard de Lyon, IPNL, IN2P3-CNRS,
     FR-69622 Villeurbanne Cedex, France
    \label{LYON}}
\titlefoot{Univ. d'Aix - Marseille II - CPP, IN2P3-CNRS,
     FR-13288 Marseille Cedex 09, France
    \label{MARSEILLE}}
\titlefoot{Dipartimento di Fisica, Universit\`a di Milano and INFN-MILANO,
     Via Celoria 16, IT-20133 Milan, Italy
    \label{MILANO}}
\titlefoot{Dipartimento di Fisica, Univ. di Milano-Bicocca and
     INFN-MILANO, Piazza delle Scienze 2, IT-20126 Milan, Italy
    \label{MILANO2}}
\titlefoot{Niels Bohr Institute, Blegdamsvej 17,
     DK-2100 Copenhagen {\O}, Denmark
    \label{NBI}}
\titlefoot{IPNP of MFF, Charles Univ., Areal MFF,
     V Holesovickach 2, CZ-180 00, Praha 8, Czech Republic
    \label{NC}}
\titlefoot{NIKHEF, Postbus 41882, NL-1009 DB
     Amsterdam, The Netherlands
    \label{NIKHEF}}
\titlefoot{National Technical University, Physics Department,
     Zografou Campus, GR-15773 Athens, Greece
    \label{NTU-ATHENS}}
\titlefoot{Physics Department, University of Oslo, Blindern,
     NO-1000 Oslo 3, Norway
    \label{OSLO}}
\titlefoot{Dpto. Fisica, Univ. Oviedo, Avda. Calvo Sotelo
     s/n, ES-33007 Oviedo, Spain
    \label{OVIEDO}}
\titlefoot{Department of Physics, University of Oxford,
     Keble Road, Oxford OX1 3RH, UK
    \label{OXFORD}}
\titlefoot{Dipartimento di Fisica, Universit\`a di Padova and
     INFN, Via Marzolo 8, IT-35131 Padua, Italy
    \label{PADOVA}}
\titlefoot{Rutherford Appleton Laboratory, Chilton, Didcot
     OX11 OQX, UK
    \label{RAL}}
\titlefoot{Dipartimento di Fisica, Universit\`a di Roma II and
     INFN, Tor Vergata, IT-00173 Rome, Italy
    \label{ROMA2}}
\titlefoot{Dipartimento di Fisica, Universit\`a di Roma III and
     INFN, Via della Vasca Navale 84, IT-00146 Rome, Italy
    \label{ROMA3}}
\titlefoot{DAPNIA/Service de Physique des Particules,
     CEA-Saclay, FR-91191 Gif-sur-Yvette Cedex, France
    \label{SACLAY}}
\titlefoot{Instituto de Fisica de Cantabria (CSIC-UC), Avda.
     los Castros s/n, ES-39006 Santander, Spain
    \label{SANTANDER}}
\titlefoot{Dipartimento di Fisica, Universit\`a degli Studi di Roma
     La Sapienza, Piazzale Aldo Moro 2, IT-00185 Rome, Italy
    \label{SAPIENZA}}
\titlefoot{Inst. for High Energy Physics, Serpukov
     P.O. Box 35, Protvino, (Moscow Region), Russian Federation
    \label{SERPUKHOV}}
\titlefoot{J. Stefan Institute, Jamova 39, SI-1000 Ljubljana, Slovenia
     and Laboratory for Astroparticle Physics,\\
     \indent~~Nova Gorica Polytechnic, Kostanjeviska 16a, SI-5000 Nova Gorica, Slovenia, \\
     \indent~~and Department of Physics, University of Ljubljana,
     SI-1000 Ljubljana, Slovenia
    \label{SLOVENIJA}}
\titlefoot{Fysikum, Stockholm University,
     Box 6730, SE-113 85 Stockholm, Sweden
    \label{STOCKHOLM}}
\titlefoot{Dipartimento di Fisica Sperimentale, Universit\`a di
     Torino and INFN, Via P. Giuria 1, IT-10125 Turin, Italy
    \label{TORINO}}
\titlefoot{Dipartimento di Fisica, Universit\`a di Trieste and
     INFN, Via A. Valerio 2, IT-34127 Trieste, Italy \\
     \indent~~and Istituto di Fisica, Universit\`a di Udine,
     IT-33100 Udine, Italy
    \label{TU}}
\titlefoot{Univ. Federal do Rio de Janeiro, C.P. 68528
     Cidade Univ., Ilha do Fund\~ao
     BR-21945-970 Rio de Janeiro, Brazil
    \label{UFRJ}}
\titlefoot{Department of Radiation Sciences, University of
     Uppsala, P.O. Box 535, SE-751 21 Uppsala, Sweden
    \label{UPPSALA}}
\titlefoot{IFIC, Valencia-CSIC, and D.F.A.M.N., U. de Valencia,
     Avda. Dr. Moliner 50, ES-46100 Burjassot (Valencia), Spain
    \label{VALENCIA}}
\titlefoot{Institut f\"ur Hochenergiephysik, \"Osterr. Akad.
     d. Wissensch., Nikolsdorfergasse 18, AT-1050 Vienna, Austria
    \label{VIENNA}}
\titlefoot{Inst. Nuclear Studies and University of Warsaw, Ul.
     Hoza 69, PL-00681 Warsaw, Poland
    \label{WARSZAWA}}
\titlefoot{Fachbereich Physik, University of Wuppertal, Postfach
     100 127, DE-42097 Wuppertal, Germany
    \label{WUPPERTAL}}
\addtolength{\textheight}{-10mm}
\addtolength{\footskip}{5mm}
\clearpage
\headsep 30.0pt
\end{titlepage}
%%%%%%%%%%%%%%%%%%%%%%%%%
%
% Change for the document body
%\pagestyle{heading} % for page numbering
\pagenumbering{arabic} % page numbering in number
\setcounter{footnote}{0} %
\large
\section{Introduction}

The measurement of the branching ratio for the decay 
$b \rightarrow u \ell \bar \nu$ provides the most precise way to determine the 
$|V_{ub}|$ element of the Cabibbo-Kobayashi-Maskawa (CKM) mixing matrix. 
Evidence for a non-zero
value of $|V_{ub}|$ was first obtained \cite{cleo1,argus1} 
by observing leptons produced in $B$ decays 
with momentum above the kinematic limit 
for $b \rightarrow c \ell \bar \nu$ transitions. 
However, extracting $|V_{ub}|$ from the yield of leptons 
above the $b \rightarrow c \ell \bar \nu$ endpoint 
is subject to a large model dependence. More recently, exclusive 
$B \rightarrow \pi \ell \bar \nu$ and $B \rightarrow \rho \ell \bar \nu$ 
decays were observed and their rates measured~\cite{cleo2,cleo3}. 
But the determination of $|V_{ub}|$ from exclusive semileptonic decays
%contrary to the case for $|V_{cb}|$ in $B \rightarrow D^* \ell \nu$, 
also has a significant model dependence.

\begin{figure}
\begin{center}
%\begin{tabular}{c c}
\epsfig{file=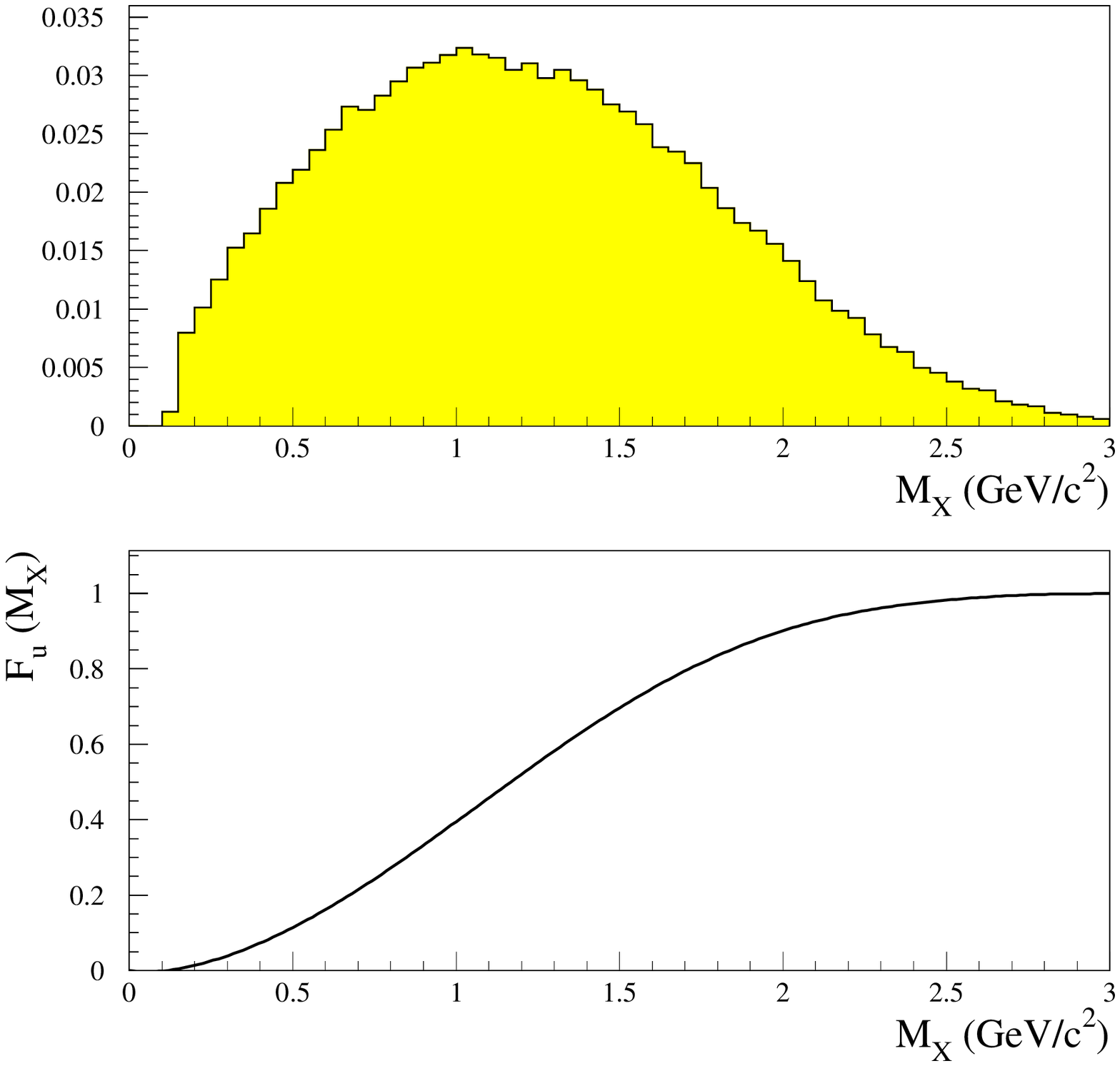,width=14.0cm,height=15.0cm}
%\end{tabular}
\end{center}
\caption[]{\sl The invariant mass spectrum $M_X$ of the $u$-spectator quark 
system for inclusive $B \rightarrow X_u \ell \bar \nu$ decays at the parton 
level obtained with a dedicated decay generator~\cite{btool} (upper plot) and 
the fraction of the decays with $M_X$ below a given value (lower plot).}
\label{fig:1}
\end{figure}

The extraction of $|V_{ub}|$ from the distribution of the invariant mass $M_X$
of the hadronic system recoiling against the lepton pair in 
$B \rightarrow X_u \ell \bar \nu$ 
transitions was proposed several years ago~\cite{barger} and it has recently
been the subject of new theoretical calculations~\cite{falk,bdu}. 
There have been two other $|V_{ub}|$ determinations at LEP based on 
inclusive analysis of semileptonic decays~\cite{lep}. 
The method used here starts from the observation that in most 
$B \rightarrow X_u \ell \bar \nu$ decays the hadronic system 
recoiling against the $\ell \bar \nu$ has an invariant mass below
the charm mass (see Fig.~\ref{fig:1}). 
Because a much larger fraction of the total rate is involved, 
the model dependence when extracting $|V_{ub}|$ from the decay rate 
to such states 
is much smaller than when using the decay rate to leptons 
above the $b \rightarrow c \ell \bar \nu$ endpoint 
or that to exclusive final states~\cite{falk,bdu}.

This paper presents the first determination of $|V_{ub}|/|V_{cb}|$ 
based mainly on candidate $B$ semileptonic decays 
with reconstructed hadronic invariant masses below the 
$D$ mass
and enriched in $b \rightarrow u$ transitions 
using the secondary vertex topology and identified kaons and protons.
%result was extracted by a fit to the 
%number of background subtracted selected events and 
The shape of the lepton energy spectrum in the $B$ rest frame
is also used. 
%This technique allowed to reduce the systematic uncertainties 
%due to the description of $b \rightarrow c$ decays. 

Section~2 describes the event preselection, 
the particle identification, the reconstruction of the hadronic secondary 
system and of the $B$ energy and direction, and the $b \rightarrow u$ 
enrichment. Section~3 presents the final event sample, the extraction 
of $|V_{ub}|/|V_{cb}|$, the stability checks, and the evaluation of the
systematic errors. Section~4 summarises. 

\section{Data Analysis}

%The extraction of the value of the $|V_{ub}|$ element is based on the analysis
%of the secondary hadronic mass and of the lepton energy, computed in the $B$ 
%rest frame, for semileptonic $B$ decays. 

The analysis was performed using data collected by the 
{\sc Delphi} detector at {\sc Lep} at centre-of-mass energies around the 
$Z^0$ pole between 1993 and 1995, corresponding to 2.8~$\times 10^6$ $Z^0$ 
hadronic decay candidates. The {\sc Delphi} detector was described in detail
in~\cite{delphidet} and its performance was reviewed in~\cite{delphiperf}.
The backgrounds were 
estimated using samples of $Z^0$ hadronic decays generated with 
{\sc Jetset 7.3}~\cite{jetset} and passed through the full detector simulation.
These simulated events corresponded to 4.9 times the data\footnote{
The numbers of simulated events quoted hereafter in this paper have all been
renormalised by this factor 4.9 so as to be directly comparable with the
numbers of data events quoted}
and were evenly divided in order to describe the {\sc Delphi} detector response
in the  different years of data taking. 
Background $B \rightarrow X_c \ell \bar \nu$ decays were 
generated using, for the exclusive modes, form factors based on a relativistic
quark model~\cite{bsw}.

Events containing signal $B \rightarrow X_u \ell \bar \nu$ decays were
simulated  using a dedicated decay generator~\cite{btool} interfaced with 
{\sc Jetset} and passed through the full {\sc Delphi} detector simulation.
Hadronic final states were produced using a tuned version of the parton
shower model. The values of the branching ratios for the exclusive 
$B \rightarrow \pi \ell \bar \nu$ and $B \rightarrow \rho \ell \bar \nu$ 
decays were forced to those measured by {\sc Cleo}~\cite{cleo2,cleo3}. 
The probability of producing vector and axial vector resonances was 
tuned to agree with the measurements of their inclusive cross sections in 
$Z^0$ decays.

\subsection{Event Preselection and Reconstruction}

Hadronic events were selected using the standard {\sc Delphi} 
criteria~\cite{hadsel}.
These yielded 2789419 events in the combined 1993-95 data. The sample
was enriched in $Z^0 \rightarrow b \bar b$ events by applying a
$b$-tag algorithm based on measurements of the track impact parameters. 
This algorithm
computes the probability for all reconstructed particles to originate from 
the event primary vertex~\cite{btag}. This probability was required to 
be smaller than 0.08 corresponding to an efficiency of 85\% for 
$Z^0 \rightarrow b \bar b$ and a rejection factor of about 7 for other 
hadronic decays. Events were divided into two hemispheres by a plane 
perpendicular to the event thrust axis. Jets were reconstructed using the
{\sc Luclus} clustering algorithm~\cite{jetset} with a $d_{join}$ value of 
6.0~GeV. Only events with two or three jets were 
used. For the two most energetic jets in each event, a secondary vertex
reconstruction in the jet was performed using those charged particle 
tracks with significantly large impact parameters. This procedure allowed to 
inclusively reconstruct a jet secondary vertex in 60\% of all the jets in 
events satisfying the $b$-tagging criteria.

\subsection{Particle Identification}

For this analysis, hadronic $b$-tagged events were required to contain 
one identified lepton ($e$ or $\mu$).
Candidate leptons from semileptonic $B$ decay were selected in the 
momentum interval 3.5~GeV/c~$< p <$~25~GeV/c.

Muons were identified by the hits associated in the muon 
chambers. The efficiency was 
estimated from simulation to be (83.0$\pm$2.0)\%. 
The probability for a hadron to be misidentified as a muon
was measured on data to be (0.68$\pm$0.03)\%.

Electron candidates were selected using a Neural Network based on the 
response of the HPC electromagnetic (e.m.) calorimeter and on the measured 
specific ionisation ($dE/dx$) in the TPC. The efficiency of this selection 
criteria was measured with Compton events in data, yielding
(70.0$\pm$2.0)\% with a misidentification probability of 
(0.38$\pm$0.03)\%.
In order to reduce the background from $b \rightarrow c \rightarrow \ell$ and
$c \rightarrow \ell$ transitions the lepton candidates were required to
have $p_t^{in} >$ 0.5~GeV/c, where $p_t^{in}$ is the momentum transverse to 
the jet axis reconstructed including the lepton candidate.
In order to ensure its accurate extrapolation to the production point, each 
lepton candidate was required to have at least one associated hit in the 
silicon Vertex Detector and a positive lifetime-signed impact parameter 
relative to the primary vertex.

The identification of 
strange mesons, which are produced in the cascade $b \rightarrow c \rightarrow
s$ decay, was used to reduce the background from charmed $b$ decays. 
Kaons and protons, with $p >$ 2.5~GeV/c, were identified by the  
combination of the response of the 
{\sc Delphi} Ring Imaging CHerenkov (RICH) detectors and the $dE/dx$ in the 
TPC~\cite{ribmean}. $K^0_s$ were reconstructed in their 
$\pi^+\pi^-$ decay mode~\cite{delphiperf}, requiring a total momentum larger 
than 2.5~GeV/$c$.

Candidate $\pi^0 \rightarrow \gamma \gamma$ decays were also tagged.
At energies up to about 6~GeV, the two photons from the $\pi^0$ decay are 
separated enough to produce two distinct electromagnetic showers in the e.m. 
calorimeter. Photon pairs with a total energy below 9~GeV and with an 
invariant mass 0.045~GeV/c$^2 < m_{\gamma \gamma} <$ 0.225~GeV/c$^2$ were
accepted as $\pi^0$ candidates. At larger energies, $\pi^0$'s were 
discriminated from photons by the e.m. cluster shape reconstructed in the HPC 
calorimeter.

\subsection{Reconstruction of the Secondary Hadronic System}

A total of 52952 hemispheres with an identified lepton were accepted in 
data and 52661 for the simulated backgrounds.

The secondary hadronic system was reconstructed using an inclusive 
method. First 
%the probability for a given particle to originate from a 
%$B$ decay product was computed from 
a likelihood variable was computed as the 
product of the likelihood ratios that the particle originated 
from the $B$ decay or from the primary hadronisation
for a set of discriminating variables. These were divided into two categories. 
The first consisted of the following six kinematical variables: transverse 
momentum $p_t$, ratio of particle momentum and jet energy $p/E_{jet}$, 
particle mass, rapidity, rank in decreasing energy order, and increase of 
the invariant mass of the particles associated to the jet secondary vertex by 
the addition of this particle. 
These variables were computed for all charged particles with momenta above 
0.5~GeV/c and neutral particles with energy above 1.0~GeV. 
The second category grouped four topological variables computed only for
charged particle tracks with $p >$~0.7~GeV/c and associated hits in the 
Vertex Detector. These variables were the track impact parameters relative to 
the primary event vertex in the $R-\phi$ and $z$ projections normalised to
their errors, the $\chi^2$ contribution of the track to the jet inclusive
secondary vertex, and the distance of the point of closest approach of the
track to the jet axis normalised to the distance between the primary and
secondary vertex. 

\begin{figure}
\begin{center}
%\begin{tabular}{c c}
\epsfig{file=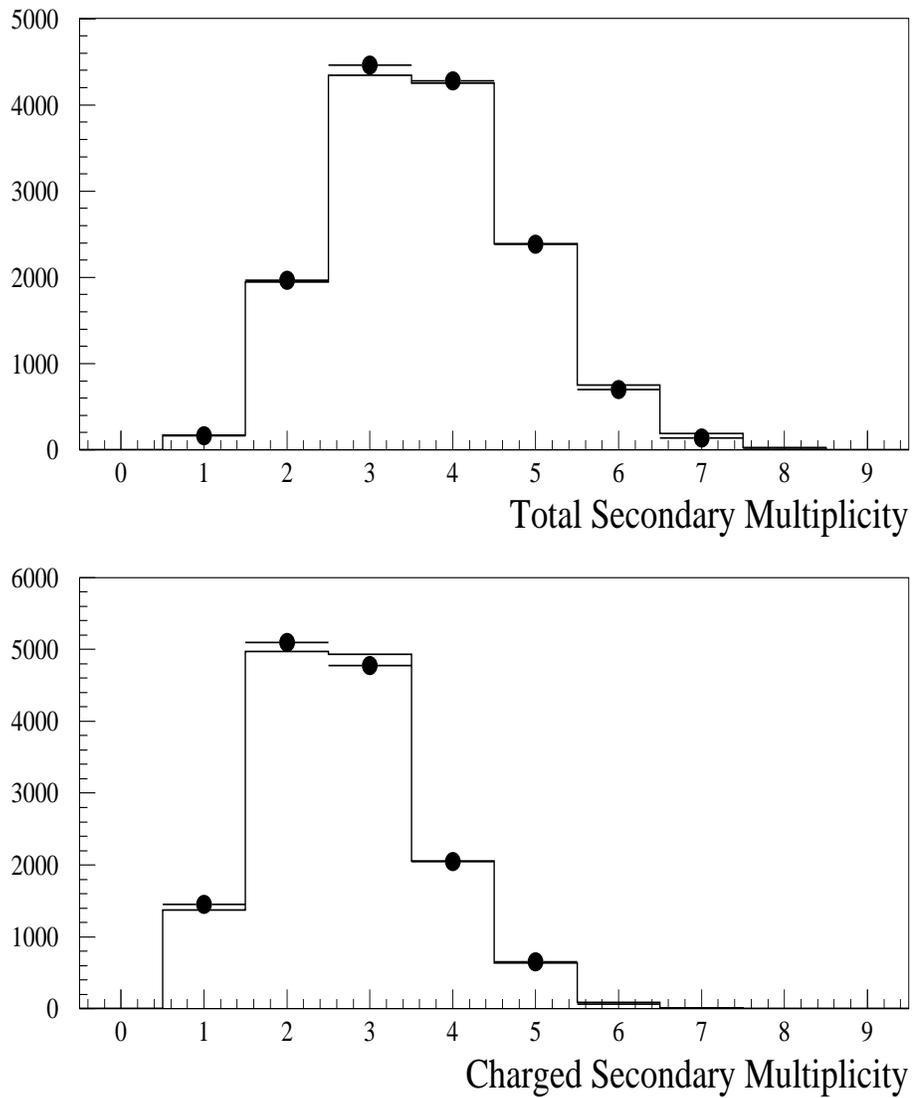,width=14.0cm,height=17.0cm}
%\end{tabular}
\end{center}
\caption[]{\sl The total (upper plot) and charged (lower plot) multiplicity of 
the reconstructed secondary hadronic system in selected decays for data 
(points with error bars) and simulation (histogram).}
\label{fig:2}
\end{figure}

\begin{figure}
\begin{center}
\epsfig{file=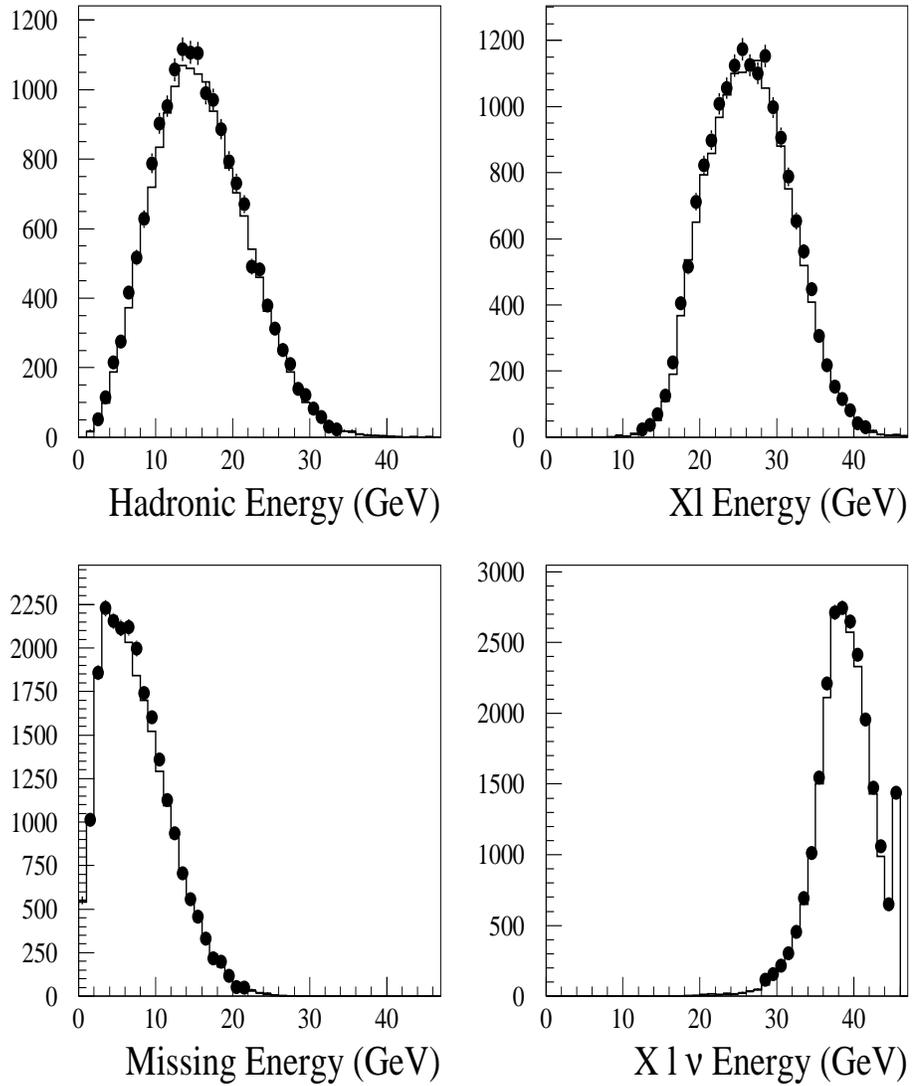,width=14.0cm,height=17.0cm}
\end{center}
\caption[]{\sl Energy Distributions: 
i) energy of the hadronic system (upper left plot),
ii) energy of the hadronic system plus the lepton (upper right plot),
iii) missing energy (lower left plot) and iv) reconstructed $B$ energy 
(lower right plot). The dots with error bars represent the data and the 
histograms the simulation.}
\label{fig:3}
\end{figure}

The hadronic secondary system was then reconstructed via an iterative 
algorithm that used the charged particles sorted by decreasing value of the 
above likelihood variable to define a secondary vertex. 
All charged particles with likelihood larger than 0.75 were tested 
for their compatibility with originating from a common secondary vertex. 
Those contributing less than 4.0 to the $\chi^2$ of the secondary vertex fit 
and giving a total invariant mass $M_X$ smaller than 3.0~GeV/c$^2$ were 
accepted. 
After the charged hadronic system was defined, identified $K^0_s$ and 
$\pi^0$ candidates with a likelihood larger than 0.65 were tested. 
At most two $\pi^0$'s were accepted at each vertex, if the total invariant mass
of the particles associated to the vertex 
did not exceed 3.0~GeV/c$^2$. Reconstructed secondary hadronic systems
with only neutral particles, with an absolute value of the charge above 1, 
or with a total energy smaller than 4~GeV were rejected. Secondary hadronic
systems consisting of a single particle were accepted if the particle was
consistent with forming a common vertex with the lepton, in this case the 
mass $M_X$ was set to the $\pi$ mass.
The total and charged multiplicities of the secondary system for 
data and simulation are shown in Fig.~\ref{fig:2}. 
The corresponding energy distributions for the secondary hadronic system $X$, 
the $X\ell$ system, the missing energy in the hemisphere, and the $X\ell\bar\nu$
system are shown in Fig.~\ref{fig:3}. 
A secondary hadronic system was reconstructed in 75\% of simulated 
$B \rightarrow X_c \ell \bar \nu$ decays
satisfying the event selection criteria described above,
and in 69\% of signal $B \rightarrow X_u \ell \bar \nu$ decays.
37986 decays were reconstructed in the data and 37899 in the simulation.

\begin{figure}
\begin{center}
%\begin{tabular}{c c}
\epsfig{file=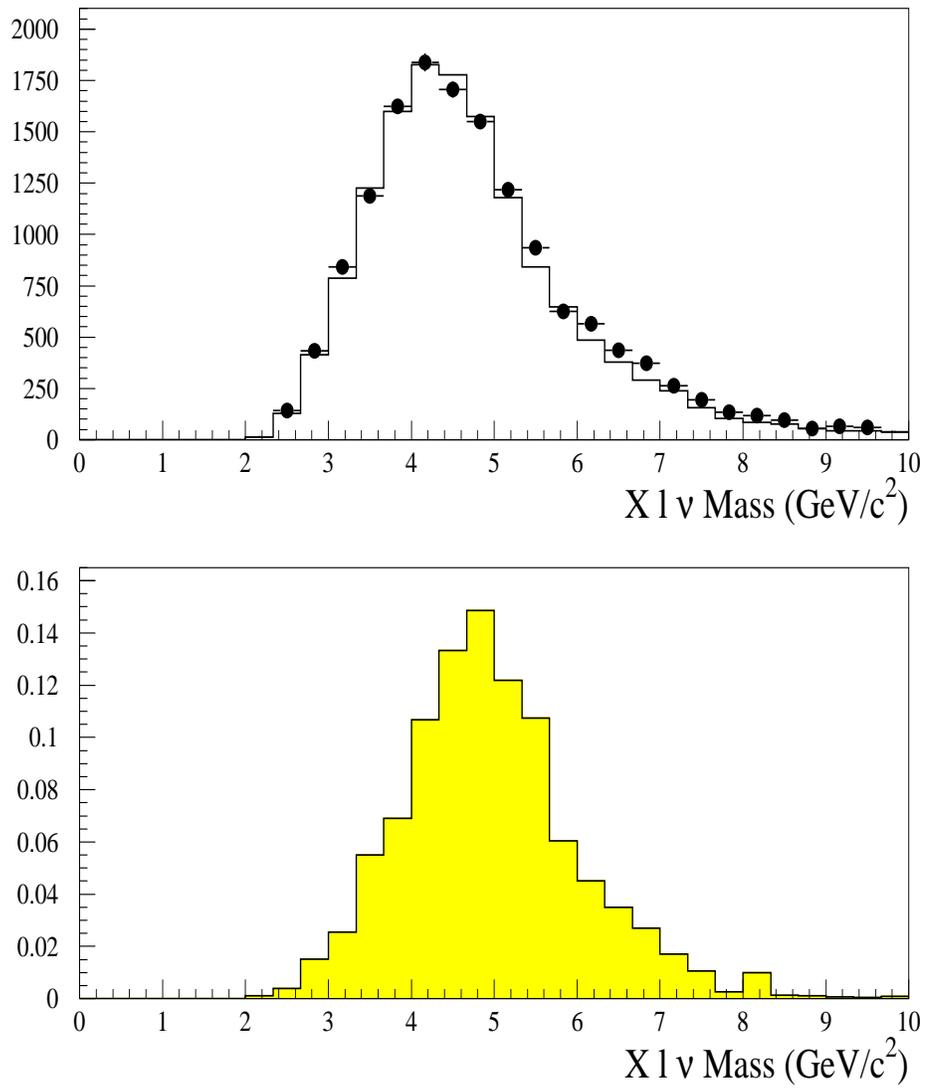,width=14.0cm,height=17.0cm}\\
%\epsfig{file=lvtx_mxln_1n.eps,width=10.0cm,height=10.0cm}\\
%\epsfig{file=lvtx_mxln_2n.eps,width=10.0cm,height=10.0cm}
%\end{tabular}
\end{center}
\caption[]{\sl Invariant mass $M_{X \ell \bar \nu}$ of the reconstructed $B$ 
decay:
data (points with error bars) and simulated background (histogram) (upper
plot) and $b \rightarrow u$ simulated signal (lower plot).}
\label{fig:4}
\end{figure}

The total mass of the candidate $B$ decay, $M_{X \ell \bar \nu}$, was 
estimated from the invariant mass of the system formed by the secondary 
hadronic system, the lepton and the neutrino as discussed below. 
For $B \rightarrow X_u \ell \bar \nu$ and fully reconstructed 
$B \rightarrow X_c \ell \bar \nu$ decays, $M_{X \ell \bar \nu}$ peaks 
at $\sim 5.0$~GeV/c$^2$ and has a resolution of 0.9~GeV/c$^2$, while for
partially reconstructed $B \rightarrow X_c \ell \bar \nu$ decays  
$M_{X \ell \bar \nu}$ peaks at 4.5~GeV/c$^2$ (see Fig.~\ref{fig:4}).
The latter decays contribute a background at values of $M_X$ below the charm 
mass. They can be identified in part because of their lower value of 
$M_{X \ell \bar \nu}$ compared with fully reconstructed decays. Therefore,  
decays with $M_{X \ell \bar \nu} <$ 3.0~GeV/c$^2$ were removed. 
For decays with  3.0~GeV/c$^2$ $< M_{X \ell \bar \nu} <$ 4.5~GeV/c$^2$ and 
$M_X <$ 1.6~GeV/c$^2$, the measured hadronic mass $M_X$ was rescaled by 
$M_B / M_{X \ell \bar \nu}$.  Of the decays in this category that 
passed the final selections (see Table~1, below), this rescaling promoted 
46.2\% to $M_X >$ 1.6~GeV/c$^2$ in data and 45.8\% in simulation.
In order to further remove partially reconstructed $D$ decays, all charged 
particles in the lepton hemisphere with $p >$ 1.5~GeV/c which were not 
associated with the secondary vertex were tested for their probability
to originate at the event primary vertex. Decays giving a probability below 
0.025 were removed, since this low probability indicates the presence of 
additional secondary particles that were not included. 
In addition, decays with two 
identified leptons in the same hemisphere were removed, because double 
semileptonic $B \rightarrow X_c \ell \bar \nu$, $X_c \rightarrow X_s \bar \ell
\nu$ decays result in a low mass hadronic system and represent a background 
to this analysis.
The secondary hadronic system reconstruction and selection gave 34583 accepted 
hemispheres in data and 33769 in simulation. 

Finally, an inclusive search for $D^* \rightarrow D \pi$ was 
performed. Charged pion candidates with 0.4~GeV/c $< p <$~3.0~GeV/c and 
$p_t <$ 0.7~GeV/c and $\pi^0$ candidates with 1.5~GeV $< E <$ 3.0~GeV and 
$E_t <$ 0.7~GeV were added in turn to the secondary system $X$ and the 
mass difference $\Delta M = M_{X \pi} - M_{X}$ was computed. Events with 
0.14~GeV/c$^2 < \Delta M <$~0.16~GeV/c$^2$ and a secondary hadronic mass 
above 0.6~GeV/c$^2$ were accepted as candidate $D^*$ decays. 
Their mass $M_X$ was then fixed to 2.01~GeV/c$^2$. 
%Of the decays in this category that passed the final selections, 
In the simulation, these criteria correctly identified 44\% 
of semileptonic decays with a $D^*$ meson and $M_X <$ 1.6~GeV/c$^2$.
This procedure promoted a further 13.8\% to $M_X >$ 1.6~GeV/c$^2$ in data, 
and 14.2\% in simulation.

\subsection{Boosted Lepton Energy}

In order to improve the separation of $B \rightarrow X_u \ell \bar \nu$ from 
$B \rightarrow X_c \ell \bar \nu$ and other background sources, the lepton 
energy
in the $B$ rest frame was determined. For each decay, the energy of the 
$B$ hadron (see Fig.~3) was estimated as the energy sum of the 
identified lepton, the secondary hadronic system and the neutrino energy. 
The neutrino energy was computed from the missing energy in the 
hemisphere corrected by a function of the $E_{X \ell}$ energy determined
from the simulation~\cite{emiss}. 
Neutrino energies in the range of 1.5~GeV $< E_{\nu} <$ 25~GeV and a minimum 
$B$ energy of 25~GeV were required. 
The resolution of the neutrino energy in $B \rightarrow X \ell \bar \nu$ 
decays was estimated to be 3.6~GeV. The resulting resolution of the $B$ 
energy was studied on simulation and found to be 9.8\% for 80\% of all
inclusive semileptonic $B$ decays and 15.4\% for the remaining decays.
The $B$ direction was taken as
the vector joining the primary vertex to the jet secondary vertex.
For those decays without a reconstructed secondary vertex, the sum of the 
momentum vectors of the hadronic system, of the lepton and of the missing 
momentum was computed. The angular resolution of the $B$ direction was 
estimated to be 3.2$^{\circ}$ for semileptonic $B$ decays. 

The lepton energy $E^*_{\ell}$ was computed in the 
frame defined by the estimated $B$ energy and direction. The resolution on 
the $E^*_{\ell}$ reconstruction was studied on simulated events and found
to be 14\% for 81\% of the selected decays. The values of the resolution 
obtained by analysing $B \rightarrow X_u \ell \bar \nu$ and 
$B \rightarrow X_c \ell \bar \nu$ events separately were found to be 
consistent.

\subsection{$b \rightarrow u$ enrichment}

A procedure was developed to select separate samples, enriched or 
depleted in $b \rightarrow u$ transitions, independently of the reconstructed 
hadronic mass. It relies on the sign of the lepton impact parameter 
relative to the secondary vertex position and on the presence of identified
kaons in the same hemisphere as the lepton. For each hemisphere, with a 
reconstructed secondary vertex, the lepton impact parameter $d^{\ell}_{sec}$
was computed relative to 
this vertex and signed using the lifetime convention, {\it i.e.}
it was signed negative if the lepton appeared to 
originate between the primary and the secondary vertex, and positive if it was
downstream of the secondary vertex. In $B \rightarrow X_c \ell \bar \nu$ 
decays, the secondary vertex corresponds mainly to the charm decay vertex. 
Consequently, 
$b \rightarrow c$ semileptonic transitions tend to give leptons with 
negatively signed impact parameters, since the lepton comes from the $B$ decay
vertex. But in $B \rightarrow X_u \ell \bar \nu$ transitions, 
the secondary vertex coincides with the $B$ decay vertex and thus with the 
lepton production point,
so the impact parameter signing depends only on 
resolution effects and is positive or negative with equal probability.
Therefore, events with a reconstructed secondary vertex and a significantly
negative lepton impact parameter were assigned to the $b \rightarrow u$
depleted class.
Requiring $d^{\ell}_{sec} < -0.015$~cm selected 29.4\% of the decays 
fulfilling the final selection in data, and 28.7\% in the simulated background 
sample. 
Those with a single secondary particle not identified as a kaon or a proton
were assigned to the $b \rightarrow u$ enriched class.
 
The detection of a strange particle in the semileptonic $B$ decay was also 
used to separate cascade $B \rightarrow X_c \ell \bar \nu$ followed by 
$X_c \rightarrow K^{\pm} X$ decays from $B \rightarrow X_u \ell \bar \nu$ 
transitions, where the production of strange particles is
suppressed because they can originate only from the spectator $s$ quark in 
$B_s$ decays or from the production of an $s \bar s$ pair in the hadronisation
process. 
%Kaons, protons and $K^0_s$ were
%identified and required to be compatible with being $B$ decay products based on
%the likelihood variable discussed in section~2.3. 
In the same hemisphere as the lepton, 
36.5\% of reconstructed decays contained an identified $K^{\pm}$, $K^0_s$ 
or proton in the data, and 37.8\% in the simulated backgrounds, 
while the simulation predicts 14\% in $b \rightarrow u$ transitions. 
These decays were assigned to the $b \rightarrow u$ depleted class 
if $d^{\ell}_{sec} <$~0. 

\section{Results}

Candidate semileptonic $B$ decays were further selected by imposing
the following selection in order to remove background and poorly reconstructed
decays. The summed energy of the hadronic system and the lepton was 
required to be larger than 12~GeV and larger than 70\% of the jet energy. 
Decays with an invariant mass of the secondary hadronic system and of the 
lepton $M_{X \ell}$ below 2.0~GeV/c$^2$ were also removed. 
Finally, decays in which the lepton charge had a
sign equal to that of the hadronic system were discarded.
These criteria selected 12134 decays in data.
The $Z^0$ simulated sample, which contained no $b \rightarrow u$ transitions, 
gave 11695 expected events, while from a dedicated signal sample an  
efficiency for $B \rightarrow X_u \ell \bar \nu$, with $\ell = e, \mu$, of 
(9.3$\pm$0.3)\% was obtained. The background composition was studied on
simulation, for $E^*_{\ell} >$~0.8~GeV, and found to consist of 90\% of 
$B \rightarrow X_c \ell \bar \nu$ decays, 8\% of cascade 
$B \rightarrow X_c \rightarrow X_s \bar \ell \nu$ decays and 
2\% of $D \rightarrow X_s \bar \ell \nu$ decays and misidentified hadrons. 

Selected decays were divided into four independent classes according 
to the reconstructed secondary hadronic mass $M_X$ and the 
$b \rightarrow u$ enrichment criteria, described in section~2.5. These are:
i) $b \rightarrow u$ enriched decays with $M_X <$ 1.6~GeV/c$^2$, 
ii) $b \rightarrow u$ enriched decays with $M_X >$ 1.6~GeV/c$^2$,
iii) $b \rightarrow u$ depleted decays with $M_X <$ 1.6~GeV/c$^2$, 
and iv) $b \rightarrow u$ depleted decays with $M_X >$ 1.6~GeV/c$^2$.
The $M_X$ value of 1.6~GeV/c$^2$ was chosen on the basis of simulation studies.
These showed that this value 
(a) was large enough to ensure a reduced model dependence in the extraction 
of $|V_{ub}|$~\cite{bdu,btool}, 
(b) was sufficiently below the $D$ mass to suppress the bulk of 
$B \rightarrow X_c\ell\bar\nu$ decays, 
and 
(c) minimised the statistical error on $|V_{ub}|/|V_{cb}|$.

The numbers of events selected in data, the numbers of expected background 
events, and the expected fractions of the total number of signal events in the 
four classes are summarised in Table~\ref{tab:sum}.
The background was rescaled by the normalisation factor obtained from 
the fit described in section 3.1 including the $\pm 0.01$ error on this 
normalisation. For decays selected in the low $M_X$ and $b \rightarrow u$ 
enriched class, which is expected to contain 68\% 
of the $b \rightarrow u$ signal, an excess of 214$\pm$56 events above the 
expected background was found in the data 
(see Figs.~\ref{fig:6},\ref{fig:7}). 
No significant excess was observed in the other classes, where the 
$\chi^2$ probability of a deviation of the data from the prediction
larger than that observed is 30\%.

As a cross-check, the analysis was repeated using both anti-$b$-tagged
events and decays with same-sign lepton and hadronic vertex combinations.
All other selection criteria were kept as in the main analysis.
Both these samples are expected to be depleted in signal $b \rightarrow u$ 
decays but they are sensitive to possible discrepancies between data and
simulation in the description of backgrounds. The numbers of selected 
decays in the low $M_X$ and $b \rightarrow u$ class were computed. 
For the anti $b$-tagged sample 32 events were observed in the 
data compared to 33$\pm$3 expected from background. 
The same-sign sample 
consisted of 340 events in data with 317$\pm$8 expected from 
backgrounds. No excess of events in data was observed in either 
of these samples. 
 
\begin{table}[h!]
\caption[]{\sl Numbers of events selected in the data, expected 
background events, and expected fractions of the total number of signal events.
The background is rescaled by the normalisation factor obtained from 
the fit described in section 3.1, including its $\pm 0.01$ error}
\begin{center}
\begin{tabular}{|c||c|c|}
\hline 
Selection & $b \rightarrow u$ enriched & $b \rightarrow u$ 
depleted \\ \hline
\mbox { } &Data $|$~~~Back.~~~$|$ Sig.&Data $|$~~~Back.~~~$|$ Sig.\\ 
\hline\hline
$M_X <$ 1.6 GeV/c$^2$ & ~2292 $|$ 2078$\pm$30 $|$ 68\% &
1081 $|$ 1118$\pm$19 $|$ ~9\% \\ \hline
$M_X >$ 1.6 GeV/c$^2$ & ~5017 $|$ 5019$\pm$60 $|$ 17\% &
3744 $|$ 3618$\pm$46 $|$ ~6\% \\ \hline
\end{tabular}
\end{center}
\label{tab:sum}
\end{table}

In order to check the interpretation of the excess of events in the low 
$M_X$ and $b \rightarrow u$ enriched sample as a signal for 
$B \rightarrow X_u \ell \bar \nu$ transitions, a search for decays into the 
$B \rightarrow \pi \ell \bar \nu$ and $B \rightarrow \rho \ell \bar \nu$ 
exclusive final states was performed. 
Decays with 1.0~GeV~$< E^*_{\ell} <$ 3.0~GeV and a
reconstructed hadronic system consisting either of a single charged particle or
of two particles with total charge $Q = 0$ or $\pm 1$ were selected. 
The lepton energy requirement further suppressed the 
non $B \rightarrow X_c \ell \bar \nu$ backgrounds.
%The $\pi$ and $\rho$ mass regions were defined as $M_X < 0.2$~GeV/c$^2$ 
%and 0.55~GeV/c$^2 < M_X <$ 1.05~GeV/c$^2$, respectively. 
%The distributions of $M_X$ for genuine $B \rightarrow \pi \ell \bar \nu$ and 
%$B \rightarrow \rho \ell \bar \nu$ signal events, 
%satisfying the selection criteria, were studied in simulation. It was 
%found that 69\% (82\%) of all the reconstructed $\pi (\rho) \ell \bar 
%\nu$ decays have values of the hadronic mass within these regions. 
The $M_X$ distribution shows an excess of events in the data compared to
the expected backgrounds in good agreement with the expectation from 
$B \rightarrow \pi \ell \bar \nu$ and $B \rightarrow \rho \ell \bar \nu$ 
exclusive final states~(Fig.~8). 
As a cross-check, the analysis was repeated for same-sign combinations 
of the lepton and the hadronic system.
This class receives signal contributions only from 
partially reconstructed decays like 
$B^+ \rightarrow \pi^+ (\pi^-) \ell^+ \nu$, where 
the $(\pi^-)$ is not reconstructed, and background events. No significant 
excess of events was observed (see Fig.~8).
%Summing 
%both mass regions the excess consists of 41$\pm$20 events. The agreement 
%between the data and the expected background is good at values of the 
%mass $M_X$ above 1.6~GeV/c$^2$ (see Table~\ref{tab:excl}).
 
\subsection{Extraction of $|V_{ub}|/|V_{cb}|$}

The numbers of events in each decay class and their $E^*_{\ell}$ distributions
were used to determine the value of $|V_{ub}|/|V_{cb}|$ by a simultaneous 
binned maximum-likelihood fit. The ratio $|V_{ub}|/|V_{cb}|$ is given by 
the ratio of $X_u \ell \bar \nu$ to $X_c \ell \bar \nu$ decays through the
relationship~\cite{uraltsev,lephfsg}:
\begin{equation}
\begin{array}{cl}
\frac{|V_{ub}|}{|V_{cb}|} =
& \frac{0.00445}{0.04110} \times 
\left ( \frac{BR(B \rightarrow X_u \ell \bar \nu_{\ell})}
{BR(B \rightarrow X_c \ell \bar \nu_{\ell})} \times 
\frac{0.105}{0.002}\right )^{1/2} \\ 
 & \times \left (1 \pm 0.055_{QCD} \pm 0.015_{m_b} \right ) \\
\end{array}
\label{eq:vub}
\end{equation}

%\begin{equation}
%\begin{array}{cc}
%\Vcb=
%& 0.0419 \left ( \frac{BR(B \rightarrow X_c \ell \overline{\nu_{\ell}})}
%{0.105} \right )^{1/2}
%\left ( \frac{1.55}{\tau_B} \right )^{1/2}\\
%  & \left ( 1 -0.012 \frac{\mu_{\pi}^2-0.5 GeV^2}{0.1 GeV^2} \right )\\
%  & \left (1 \pm 0.015_{pert.} \pm 0.010_{m_b} \pm 0.012  \right ) \\
%\end{array}
%\label{eq:vcb}
%\end{equation}

In the fit, the overall data to simulation normalisation and the value of 
$|V_{ub}|/|V_{cb}|$ were left free to vary while the non 
$B \rightarrow X_c \ell \nu$ backgrounds were kept fixed to the fractions 
predicted from the simulation. This significantly reduced 
the systematic uncertainties from  the lepton identification and other 
sources by absorbing their effects in this overall normalisation factor. 

The result of the fit (see Fig.~7) was:
\begin{center}
${|V_{ub}|}/{|V_{cb}|}$ = 0.103 $^{+0.011}_{-0.012}$ (stat.)
\end{center} 
with the normalisation factor 1.013$\pm$0.011. 

\subsection{Stability checks}

The stability of the fitted value of $|V_{ub}|/|V_{cb}|$ 
was checked in various ways.
Repeating the analysis for electrons and muons separately gave 
0.095$\pm$0.017 and 0.107$\pm$0.014 respectively.
Moving the cut on $M_X$ from 1.6 ~GeV/c$^2$ 
% new results over a wider M_X cut range :
% > 0.098 +/- 0.013 for M_X cut at 1.05 with S/B = 0.15
% > 0.105 +/- 0.013 for M_X cut at 1.90 with S/B = 0.055
to 1.05~GeV/c$^2$ or 1.90~GeV/c$^2$, which changed
the signal-to-background ratio from 0.10 to 0.15 or 0.055,
changed the fit result to ${|V_{ub}|}/{|V_{cb}|}$ = 
0.098$\pm$0.013 or 0.105$\pm$0.013, respectively.
Several other selection criteria were also varied or dropped,
including the cuts on event $b$-tagging, lepton $p_t$, $E_{X\ell}$, 
$M_{X\ell}$, $E_B$ and the secondary vertex decay distance significance,
and the results were found to agree within the errors. 
Removing the scaling of the hadronic invariant mass $M_X$, 
described in section 2.3, gave 0.105$\pm$0.013. 
Excluding the information on the lepton energy gave 0.106$\pm$0.015. 
Keeping the simulation to data normalisation fixed gave 
$|V_{ub}|/|V_{cb}|$ = 0.108$\pm$0.009.

Finally, the analysis was repeated with an improved 
rejection of double semileptonic decays $B \rightarrow X_c \ell \bar \nu$,
$X_c \rightarrow X_s \bar \ell \nu$. Because of the presence of an additional 
neutrino and the lower secondary charged particle multiplicity, these decays 
can be confused with $B \rightarrow X_u \ell \bar \nu$ transitions. 
A loose $e$ and $\mu$ identification procedure, 
based on the $dE/dx$ measured in the TPC, 
the measured ratio of calorimetric energy to momentum, $E/p$, 
and the hit pattern in the Hadron Calorimeter, was developed. 
This had an efficiency of 75\% for leptons not already identified by the 
standard tagging procedure and a hadron misidentification probability of 18\%
in the momentum range 3~GeV/c~$< p <$~10~GeV/c.
Candidate decays were rejected if the secondary charged particle 
multiplicity (not including the seed high $p_t$ lepton) did not exceed two 
and at least one secondary particle was tagged as a lepton with these looser 
requirements.  
This gave ${|V_{ub}|}/{|V_{cb}|}$ = 0.100$\pm$0.013.

\subsection{Systematic Uncertainties}

Four categories of systematic uncertainties were considered. 
The first three affect primarily the estimation of the large $b\to c$
backgrounds.
The fourth affects primarily the evaluation of $|V_{ub}|/|V_{cb}|$ from
the observed excess in the $b\to u$-enriched sample.
The results for $|V_{ub}|/|V_{cb}|$ are summarised in Table~\ref{tab:syst}. 

\subsubsection{ Systematic errors from charm decays}
The description of charm decays affects the fraction of $B \rightarrow X_c 
\ell \bar \nu$ transitions that were accepted in the $b \rightarrow u$ 
enriched sample. 
First, the branching fractions for $D$ decays into final states 
with low charged multiplicity were considered. 
These are $D \rightarrow K^0 X$,
which in the simulation contributes 40\% of the background 
from $b\to c$ decays in the $b \rightarrow u$ enriched sample,
and $D$ decays in 0 and 1~prong final states, which contribute 46\%. 
Varying their branching ratios 
within the uncertainties of the {\sc Mark~III} measurement \cite{markiii}
changed the fitted value of $|V_{ub}|/|V_{cb}|$ by $\pm0.0062$ and $\pm0.0025$,
respectively. 
Varying the ratio of 
prompt $B$ semileptonic decays to cascade and charm decays by the 
uncertainty on the ratio BR($b \rightarrow \ell$)/BR($c \rightarrow \ell$) = 
(0.106$\pm$0.002)/(0.098$\pm$0.003)~\cite{ewwg} contributed an
uncertainty of $\pm0.0042$. 
%Summing these contributions in quadrature, the systematic uncertainties from
%the charm decay were found to be 0.0079.  

\subsubsection{ Uncertainties in $B$ hadron production and decay}
The first component is due to the fraction of
$B_s$ and beauty baryons produced. Due to the rejection of kaons and protons 
associated with the lepton hemisphere 
%and to the shorter $\Lambda_b$ lifetime, 
these beauty hadrons do not significantly contribute to the $b \rightarrow u$ 
enriched sample. 
%The production rates of $B_u$, $B_d$ and $B_s$ mesons were 
%evaluated in reference. 
Propagating the uncertainty of $\pm$~0.020 
on the sum of $B_u$ and $B_d$ meson fractions~\cite{lephfsg} 
contributed 0.0039. 
Varying the $\epsilon_b$ parameter in the Peterson $b$ 
fragmentation function according to the uncertainty in
the fraction of the beam energy taken by the beauty hadron, 
$<x_b>$ = 0.702$\pm$0.008, contributed 0.0010.
Varying the inclusive $b$ lifetime by the uncertainty of the 
present world average, (1.564$\pm$0.014)~ps~\cite{lephfsg}, 
contributed 0.0011.
The branching fraction for inclusive double charm production in $B$ decays was
fixed to 0.15 and varied by $\pm 0.03$. This changed the 
fit result by $\pm 0.0025$. Finally, the  dependence on the production rate 
of $D^*$ and $D^{**}$ mesons (where $D^{**}$ denotes either a non-resonant 
$D^* \pi$ final state or a $D^{(*)}_J$ higher excited charmed 
meson state) in $B$ semileptonic decays was studied. These states flip 
the charge of the resulting $D$ meson and thus increase the charged 
multiplicity in $b$ hadronic decays. Using the values
BR($B \rightarrow D^* \ell \bar \nu$) = (0.046$\pm$0.003)~\cite{pdg} and
BR($B \rightarrow D^{**} \ell \bar \nu$) = 
(0.034$\pm$0.006)~\cite{delphid}, gave a sum of 0.080$\pm$0.007, 
corresponding to a contribution to the systematic error of 0.0033. 
In addition, the amount of non-resonant $D^{(*)} \pi$ states was varied 
from zero to 50\% of the $D^{**}$ yield. This contributes systematic 
uncertainties both from the shape of the lepton energy spectrum and from the 
vertex topology and secondary charged multiplicity corresponding to an 
uncertainty of $\pm 0.0065$. 

Finally the systematic error from the model of the shape of the lepton spectrum
in the background $b \rightarrow c$ semi-leptonic transitions was estimated. 
The lepton spectra observed for the three signal-depleted classes were found 
to agree with those from the simulation (see Fig.~6). 
In addition, a sample enriched in $B \rightarrow D^* (X) \ell \bar \nu$ decays
was compared with the simulation prediction 
and was also found to be in good agreement.
The spectrum predicted by the {\sc Delphi} simulation program 
was compared with that from the ISGW-2~model~\cite{isgw2} 
implemented in the {\sc EvtGen} decay generator~\cite{evtgen}. 
The relative contributions of the different charm states 
were set to the central values discussed above 
in order to be sensitive only to the difference 
in the predicted shape of the lepton spectrum.
The difference of 0.0020 in the fit result 
was taken as the corresponding contribution to the systematic uncertainty. 
%Summing these contributions in quadrature, the systematic 
%uncertainties from $B$ production and decays were found to be 0.0090.  

\subsubsection{ Detector dependent systematics}
The first source is due to the efficiency and purity of 
the lepton identification. Efficiency and misidentification probability for
muon and electron tagging were extracted from both the simulation and
data as discussed in 2.2. The central values for the simulation were
changed within these errors and the corresponding changes on 
$|V_{ub}|/|V_{cb}|$ were found to be 0.0015 and 0.0020, respectively.

The second component of detector systematics is due to the hadronic mass 
and rest-frame lepton energy resolution. The former depends on both the 
secondary hadronic multiplicity and the single particle energy resolution.
The secondary hadronic multiplicity was studied on the 
$b \rightarrow u$ depleted sample to avoid possible biases from the presence of
signal events in the data. The total (charged) multiplicity of the 
secondary hadronic system 
was measured to be 3.64$\pm$0.01 (2.74$\pm$0.01) in the data 
and 3.63$\pm$0.01 (2.72$\pm$0.01) in the simulation. 
The systematic uncertainty was evaluated by considering a change of 
$\pm 0.02$ units of multiplicity. This gave
a variation of $|V_{ub}|/|V_{cb}|$ of $\pm$0.0065.
The component due to the particle energy resolution is dominated by the 
resolution of neutral particles. The systematic effects were checked by 
decreasing by 2\% the resolution on $M_X$ for those decays with neutral 
particles. This range was defined by a $\chi^2$ analysis of the hadronic
mass in real and simulated $b \rightarrow u$ depleted decays. It also 
corresponds to the effect from the electromagnetic energy resolution 
typically measured in $Z^0$ decays. 
The corresponding systematic uncertainty was found to be 0.0010.
%resulting in a total systematic from the hadronic mass resolution of 0.0066. 
The resolution on the neutrino energy reconstruction was varied by 
10\% and the effect was propagated to the $E^*_{\ell}$ resolution,
giving a systematic error contribution of 0.0010. 

Thirdly, the possible systematics in the decay classification were studied.
These systematics depend both on the measurement and sign of the lepton 
impact parameter relative to the secondary vertex and on the kaon
identification.
The lepton impact parameter systematic has two components.
The first is due to the lepton extrapolation and the second to the secondary 
vertex reconstruction. The effect of changing the resolution on the lepton 
track extrapolation was computed by smearing the resolution on the lepton 
impact parameter by 5\%, which corresponds to the maximum discrepancy observed 
in the resolution functions obtained in data and in 
simulation~\cite{btag2}.  
The component due to the secondary vertex position was evaluated by 
smearing the resolution on its decay length by 50~$\mu$m, which corresponds to
the additional smearing that increases by one the $\chi^2$ of a 
data-to-simulation comparison of the decay distance distributions. Summing 
these two effects in quadrature gave a systematic uncertainty of 
$\pm$0.0070.
The kaon tagging efficiency was varied by $\pm 2.5\%$, corresponding to 
the largest observed deviation of the performance of the hadron identification 
tagging in data and simulation, and the corresponding uncertainty on the 
result of the fit was 0.0025. 

Finally, the statistical error on the efficiency for selecting signal events 
contributes $\pm 0.0015$ to the systematic uncertainty.

%Summing these contributions in quadrature, the 
%detector systematics were found to be 0.0103.  

\begin{table}[ht!]
\caption[]{\sl Summary of the systematic uncertainties on $|V_{ub}|/|V_{cb}|$.}
\begin{center}
\begin{tabular}{|l|c|c|}
\hline
Source & Value $\pm $ Range & Syst. Error \\
\hline \hline
BR($D \rightarrow K^0 X$) & $0.53 \pm 0.05$ & 0.0062 \\
BR($D \rightarrow 0, 1$~prong) & $0.22 \pm 0.02$ & 0.0025 \\
B($b \rightarrow \ell$)/BR($c \rightarrow \ell$) &  
$\frac{ 0.106 \pm 0.002}{0.098 \pm 0.003}$ & 0.0042 \\ \hline
Charm decay sub-total & & 0.0079 \\ \hline\hline
$f(B_u) + f(B_d)$ &
$0.802 \pm 0.020$ & 0.0039 \\
$<x_b>$ &
$0.702 \pm 0.008$ & 0.0010 \\ 
$b$ lifetime &
$1.564 \pm 0.014$ & 0.0011 \\ 
BR($b \rightarrow c \bar c s$) &
$0.15 \pm 0.03$ & 0.0025 \\
BR($B \rightarrow D^* \ell \bar \nu + D^{**} \ell \bar \nu$) &
$0.080 \pm 0.007$ & 0.0033 \\ 
BR($B \rightarrow D^{(*)} \pi \ell \bar \nu$/BR($B \rightarrow D^{**} 
\ell \bar \nu$) &
$0.25 \pm 0.25$ & 0.0065 \\
Lepton spectrum shape & \mbox { } & 0.0020 \\ \hline
B production and decay sub-total & & 0.0090 \\ \hline\hline
$e$ / $\mu$ id. Efficiency &
$\pm 2.5\%$ & 0.0015 \\
$e$ / $\mu$ id. Purity &
$\pm 10\%$ / $\pm 4.5\%$ & 0.0020 \\
Hadronic Multiplicity &
\mbox{ } & 0.0065 \\
Neutral Energy Resolution &
\mbox{ } & 0.0010 \\
Missing Energy Resolution &
\mbox{ } & 0.0010 \\
$\ell $ Impact Parameter &
\mbox{ } & 0.0070 \\ 
$K$ id. Efficiency &
$\pm 2.5\%$ & 0.0025 \\
Signal Efficiency &
$\pm$ 3.0\% & 0.0015 \\ \hline
Detector-dependent sub-total & & 0.0104 \\ \hline\hline
$m_b$ &
$4.82 \pm 0.10$ & 0.0047 \\
$<p_b^2>$           & $0.5 \pm 0.1$ & 0.0028 \\
$b$ Kinematic Model & \mbox{ } & 0.0025 \\
Hadronisation Model & \mbox{ } & 0.0060 \\ 
QCD corrections     & \mbox{ } & 0.0050 \\ \hline
$b\to u$ model sub-total & & 0.0099 \\ \hline\hline
{\bf Total} & \mbox{ } & {\bf 0.0186} \\
\hline
\end{tabular}
\end{center}
\label{tab:syst}
\end{table}

\subsubsection{ Uncertainties in the $B \rightarrow X_u \ell \bar \nu$ model}

The predicted shape of the invariant-mass distribution in the 
$B \rightarrow X_u \ell \bar \nu$ decay, depends mainly on the kinematics of 
the heavy and spectator quarks inside the $B$ hadron and on the $b$ quark
mass. Further, the hadronisation process, transforming the $u \bar q$ system 
into the observable hadronic final state, represents an additional source of 
model uncertainties. 
These uncertainties were studied using a dedicated generator that implements 
different prescriptions for the initial state kinematics and the resonance 
decomposition of the hadronic final states \cite{btool}. 

%The first source is due to the values of the $b$ quark pole mass $m_b$ and of 
%the expectation value of the kinetic energy operator. 
Varying the $b$ quark mass by $\pm 100$~MeV/$c^2$~\cite{uraltsev,hoang,beneke},
gave a systematic error of $\pm 0.0045$. 
The value of the $b$ quark mass introduces also an uncertainty 
in the extraction of $|V_{ub}|/|V_{cb}|$ from the observed $B \rightarrow X_u 
\ell \bar \nu$ rate~\cite{uraltsev,hoang}, see Eq.~1. 
This gives a total error contribution of 0.0047.  
The average kinetic energy of the $b$~quark in the hadron, $<p_b^2>$, 
has been evaluated both from theory and from fits to experimental data. 
Results are scheme or model dependent and depend on the method used 
in their derivation, but point to the value 
of the parameter $\mu_{\pi}^2$ = (0.5$\pm$0.1) ~GeV$^2$.
This variation contributes 
$\pm 0.0015$ from the uncertainty on the hadronic mass spectrum and 
$\pm 0.0024$ from the derivation of the ratio $|V_{ub}|/|V_{cb}|$,
giving a total of $\pm 0.0028$.
%A second source of systematic effect is due to the ansatz for the 
The description of the motion of the $b$ quark inside the heavy hadron
also contributes uncertainties. 
The momentum distributions from the ACCMM model~\cite{accmm}, 
a shape function~\cite{dsu,neubert} parametrised as
$f(z) = z^a (1 - c z) e^{- c z}$, and the parton 
model~\cite{parton} were compared. For the ACCM, the $b$ quark pole mass was 
kept fixed at its central value of $m_b =$~4.82~GeV/c$^2$ and the $p_F$ value 
and the $a$ and $c$ coefficients in the QCD structure function were 
chosen to reproduce the same value of $<p_b^2>$.
For the parton model, the Peterson form~\cite{peterson} of the fragmentation 
function was adopted with $\epsilon_b$ = 0.0040. A systematic uncertainty
of $\pm 0.0025$ was evaluated. 

The production
of the hadronic final states from the $u \bar q$ pair was simulated
according to both the ISGW-2~\cite{isgw2} exclusive and a fully inclusive model
based on parton shower fragmentation~\cite{jetset}. The ISGW-2 model 
approximates the inclusive $B \rightarrow X_u \ell \bar \nu$ decay width by 
the sum over resonant final states, taking into account leading corrections 
to the heavy quark symmetry limit.
The predicted branching ratios for the different resonant final states 
were used to define the hadronic system emitted with the lepton. Another 
approach is to assume that, at sufficiently large recoil $u$ quark energies, 
the $u \bar q$ system moves away fast enough to resemble the evolution of a 
jet initiated by a light quark $q$ in $e^+e^- \rightarrow q \bar q$ 
annihilation. This was simulated by first arranging the $u \bar q$ 
system in a string configuration and then requiring it to fragment according 
to the parton shower model. Due to the extreme assumptions of the two 
models adopted, the resulting difference of 0.0160 in the fitted value of 
$|V_{ub}|/|V_{cb}|$ was assumed to correspond to a 90\% confidence region
and the $\pm 1 \sigma_{syst}$ was estimated to be $\pm 0.0060$.
%This study has quantified the systematic uncertainties from
%the $B \rightarrow X_u \ell \bar \nu$ decay model to be 0.0100.  

Additional sources of theoretical systematics, arising from the perturbative
part of the evaluation and the contribution of non perturbative corrections of
order $1/m_b^3$ contribute a $\pm 0.0050$ systematic error~\cite{uraltsev}.

\section{Summary and Discussion}

The value of the ratio $|V_{ub}|/|V_{cb}|$ was measured using a novel technique.
The technique uses the reconstructed mass $M_X$ of the secondary hadronic 
system produced in association with an identified lepton in the semi-leptonic
decay of a $B$ hadron and the rest-frame energy spectrum of that lepton. 
The $b\to u$ signal is enriched using identified kaons and protons 
and the lepton impact parameter with respect to the secondary vertex. 
The result obtained is 
\begin{center}
$|V_{ub}|/|V_{cb}|$ = 0.103 $^{+0.011}_{-0.012}$ (stat.)$\pm$0.016 (syst.) 
$\pm$ 0.010~(model).
\end{center}
Here the systematic error quoted combines the charm decay, B production and
decay, and detector-dependent systematics evaluated above, which all primarily 
affect the estimation of the $b\to c$ backgrounds, and the $b\to u$ model 
uncertainty is quoted separately. The technique adopted in this analysis
allowed both sources of systematic uncertainty to be reduced. At the critical
points of the analysis, the behaviour of the data agrees well with the 
expectations from the simulation. The result is found to be stable with respect
to variations in the analysis procedure.

There remains a possible further model dependence arising from a biased 
sampling of the decay phase space in $B \rightarrow X_u \ell \bar \nu$
transitions. 
The stability of the result when the $M_X$ cut was moved from 1.6 GeV/c$^2$
down to 1.05~GeV/c$^2$ or up to 1.90~GeV/c$^2$ (section 3.2) argues against 
this. To further check this possibility, 
the relative weights of different regions in the $M_X - E^*_{\ell}$
plane were analysed. 
Four regions of the $M_X - E^*_{\ell}$ plane were defined by selecting decays
with  $M_X$ and $E^*_{\ell}$ above and below 0.8~GeV/c$^2$ and 1.75~GeV,
respectively. The fit was repeated separately for these four regions.
%The upper $M_X$ cut was kept at 1.6~GeV/c$^2$.
The statistical weights of the four regions are given in Table~\ref{tab:wei}.
The result agrees with the expectation of a higher 
contribution from charmless semileptonic $B$ decays in the low-mass, 
high-energy and high-mass, low-energy regions and indicates no 
strong bias in the weighting of the decay phase space.

\begin{table}[h!]
\caption[]{\sl The weights of the four regions in the $M_X$-$E^{*}_{\ell}$
plane, used to check the phase space sampling in the determination of 
$|V_{ub}|/|V_{cb}|$.}
\begin{center}
\begin{tabular}{|c|c|c|}
\hline
\mbox{ } & 0.1$<M_X<$0.8~GeV/$c^2$ & 0.8$<M_X<$1.6~GeV/$c^2$ \\
\hline \hline 
1.75$<E^{*}_{\ell}<$3.0~GeV  & 0.33 & 0.21 \\ \hline
0.1$<E^{*}_{\ell}<$1.75~GeV  & 0.15 & 0.29 \\
\hline
\end{tabular}
\end{center}
\label{tab:wei}
\end{table}

While this analysis extracted the ratio of CKM elements $|V_{ub}|/|V_{cb}|$
from the fitted fraction of candidate $B \rightarrow X_u \ell \bar \nu$ decays,
it is also interesting to extract explicitly the the charmless semileptonic 
branching ratio. This was obtained from the fitted result, 
assuming $|V_{cb}|$ = (38.4$\pm$3.3)~$\times 10^{-3}$ and
$\tau_b$ = (1.564$\pm$0.014)~ps~\cite{lephfsg}. 
The result was:
\begin{center}
BR($B \rightarrow X_u \ell \bar \nu$) = 
(1.57$\pm$0.35 (stat.)$\pm$0.48 (syst.)$\pm$0.20 ($|V_{cb}|$) 
$\pm$ 0.01 ($\tau_b$) $\pm$~0.27~(model))~$\times 10^{-3} 
\times (|V_{cb}|/0.0384)^2$.
\end{center}
where the contribution of correlated model systematics in the derivation of 
$|V_{ub}|$ and $|V_{cb}|$ was taken into account.

\newpage

\noindent
{\Large \bf Acknowledgments}

\vspace{0.5cm}

\noindent
We would like to thank I.~Bigi, N.~Uraltsev and M.~Neubert for discussions on 
the modelling of the $B \rightarrow X_u \ell \bar \nu$ decays and on the 
related systematics in the extraction of $|V_{ub}|/|V_{cb}|$ and D.~Lange and 
A.~Ryd for providing an implementation of the ISGW-2 model in the simulation of
semileptonic $B$ decays.\\
We are greatly indebted to our technical 
collaborators, to the members of the CERN-SL Division for the excellent 
performance of the LEP collider, and to the funding agencies for their
support in building and operating the DELPHI detector.\\
We acknowledge in particular the support of \\
Austrian Federal Ministry of Science and Traffics, GZ 616.364/2-III/2a/98, \\
FNRS--FWO, Belgium,  \\
FINEP, CNPq, CAPES, FUJB and FAPERJ, Brazil, \\
Czech Ministry of Industry and Trade, GA CR 202/96/0450 and GA AVCR A1010521,\\
Danish Natural Research Council, \\
Commission of the European Communities (DG XII), \\
Direction des Sciences de la Mati$\grave{\mbox{\rm e}}$re, CEA, France, \\
Bundesministerium f$\ddot{\mbox{\rm u}}$r Bildung, Wissenschaft, Forschung 
und Technologie, Germany,\\
General Secretariat for Research and Technology, Greece, \\
National Science Foundation (NWO) and Foundation for Research on Matter (FOM),
The Netherlands, \\
Norwegian Research Council,  \\
State Committee for Scientific Research, Poland, 2P03B06015, 2P03B1116 and
SPUB/P03/178/98, \\
JNICT--Junta Nacional de Investiga\c{c}\~{a}o Cient\'{\i}fica 
e Tecnol$\acute{\mbox{\rm o}}$gica, Portugal, \\
Vedecka grantova agentura MS SR, Slovakia, Nr. 95/5195/134, \\
Ministry of Science and Technology of the Republic of Slovenia, \\
CICYT, Spain, AEN96--1661 and AEN96-1681,  \\
The Swedish Natural Science Research Council,      \\
Particle Physics and Astronomy Research Council, UK, \\
Department of Energy, USA, DE--FG02--94ER40817. \\
%=========================================================================%

%\newpage
%

\newpage

\newpage

\newpage

\newpage
\begin{figure}
\begin{center}
\epsfig{file=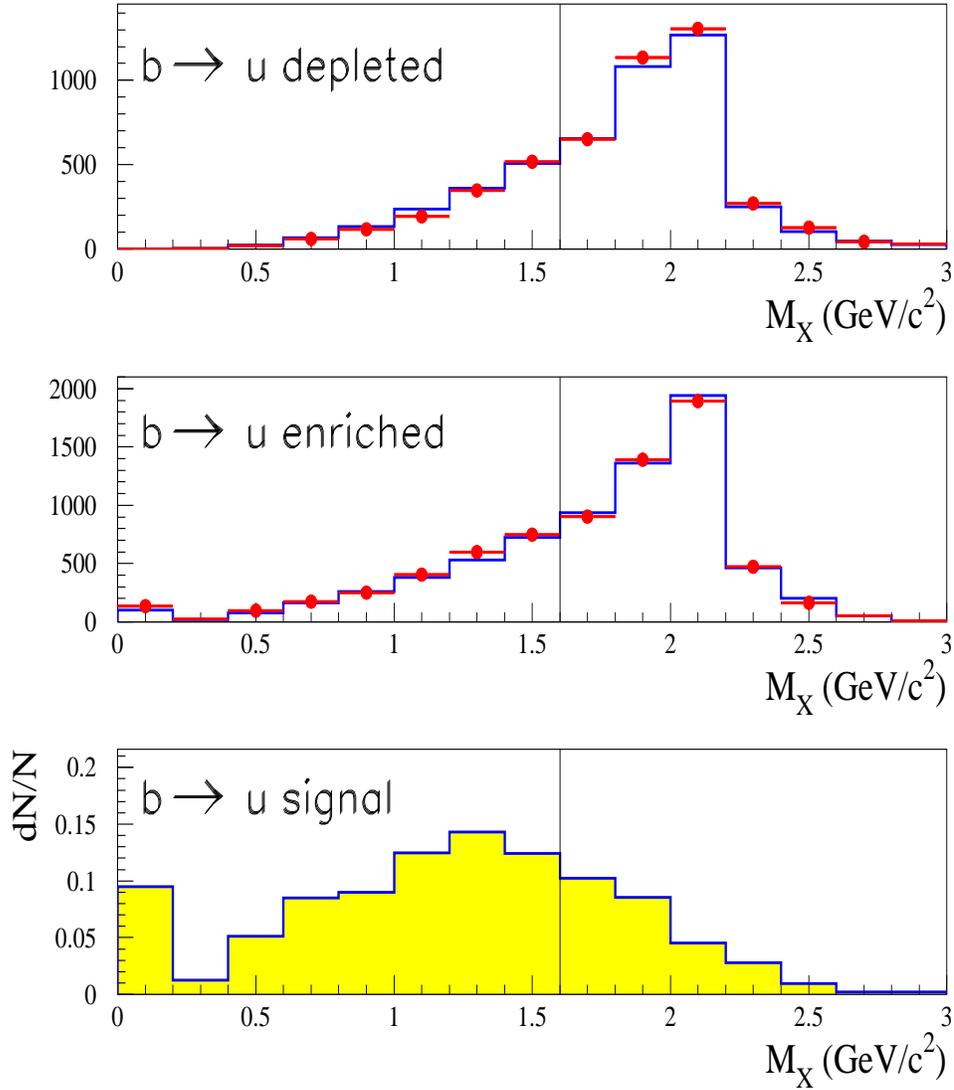,width=14.0cm,height=17.0cm}
\end{center}
\caption[]{\sl Invariant mass $M_X$ of the reconstructed secondary hadronic 
system in selected decays for data (points with error bars) and 
simulation (histogram). The plots show the $b \rightarrow u$ depleted
sample (upper plot), the $b \rightarrow u$ enriched sample (medium plot)
and the $b \rightarrow u$ signal (lower plot). The vertical lines correspond 
to the value chosen for the low $M_X$ selection.}
\label{fig:5}
\end{figure}

\newpage
\begin{figure}
\begin{center}
\epsfig{file=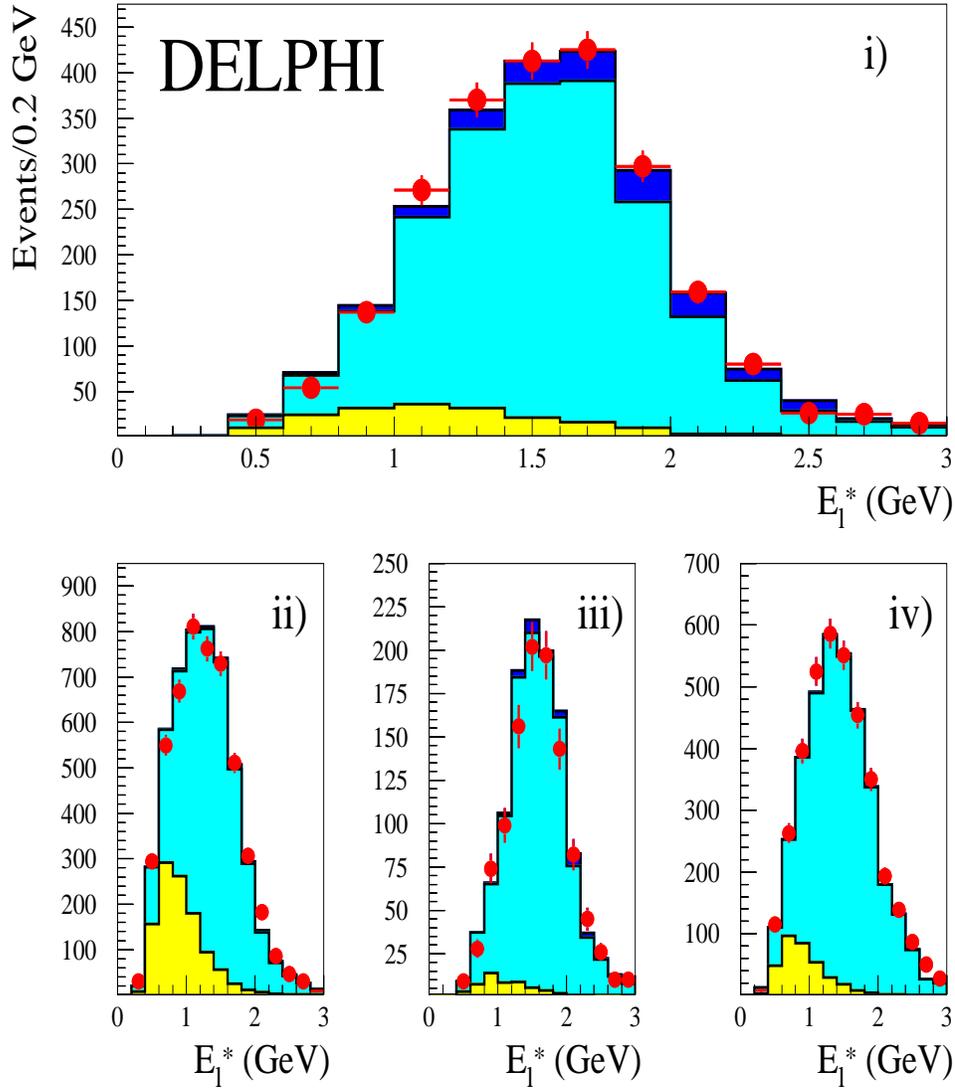,width=14.0cm,height=17.0cm}
\end{center}
\caption[]{\sl The $E^*_{\ell}$ distribution for the decays in the 
four selected classes: 
i) $b \rightarrow u$ enriched decays with $M_X <$ 1.6~GeV/c$^2$ (upper plot), 
ii) $b \rightarrow u$ enriched decays with $M_X >$ 1.6~GeV/c$^2$ (lower left),
iii) $b \rightarrow u$ depleted decays with $M_X <$ 1.6~GeV/c$^2$ (lower 
central plot), 
and iv) $b \rightarrow u$ depleted decays with $M_X >$ 1.6~GeV/c$^2$ (lower
right plot).
Data are indicated by the points with error bars, the 
$b \rightarrow X_u \ell \bar \nu$ signal by the dark shaded histograms,
the $b \rightarrow X_c \ell \bar \nu$ background by the medium shaded 
histograms, and the other backgrounds by the light shaded histograms.}
\label{fig:6}
\end{figure}

\newpage
\begin{figure}
\begin{center}
\epsfig{file=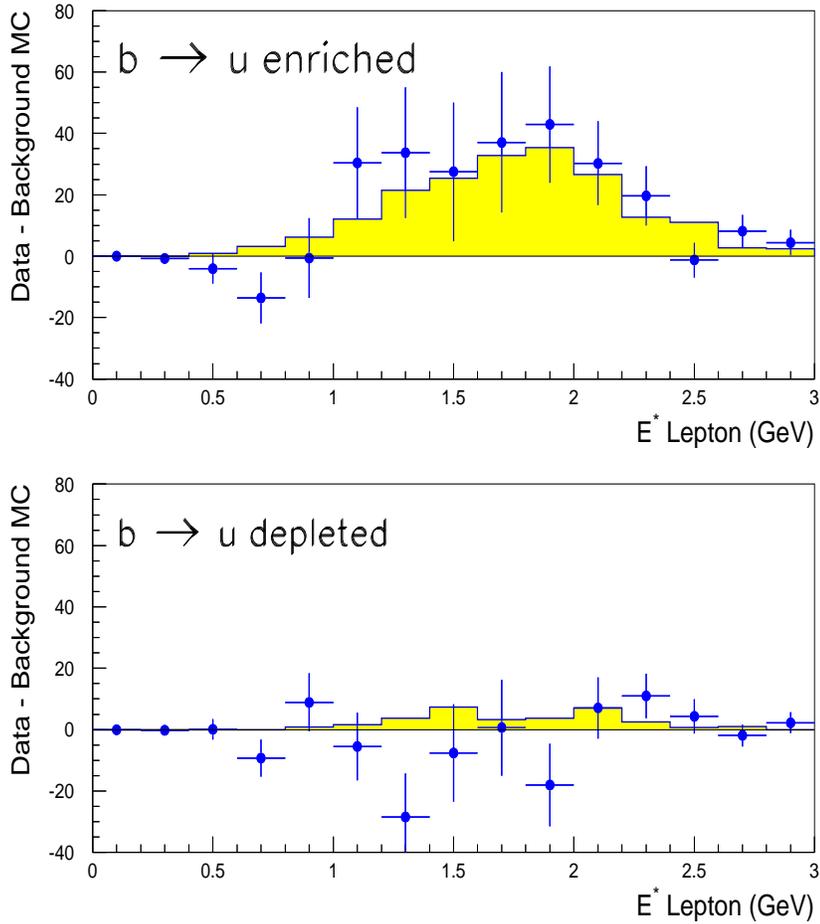,width=12.0cm,height=14.0cm}
\end{center}
\caption[]{\sl Background subtracted $E^*_{\ell}$ distributions: the 
$b \rightarrow u$ enriched decays with $M_X <$ 1.6~GeV/c$^2$ (upper plot)
and $b \rightarrow u$ depleted decays with $M_X <$ 1.6~GeV/c$^2$ (lower plot).
The  background was rescaled by the fitted normalisation factor. The 
shaded histograms show the expected $E^*_{\ell}$ distribution for signal 
$B \rightarrow X_u \ell \bar \nu$ decays normalised to the amount of signal
corresponding to the fitted $|V_{ub}|/|V_{cb}|$ value.}
\label{fig:7}
\end{figure}

\newpage
\begin{figure}
\begin{center}
\begin{tabular}{l r}
{\huge \bf $Q_{vtx}^{\pm , 0}~~Q_{\ell}^{\mp}$}~~~~~~~~~~~~~ & 
~~~~~~~~~~~~~{\huge \bf $Q_{vtx}^{\pm}~~Q_{\ell}^{\pm}$}\\
\end{tabular}
\epsfig{file=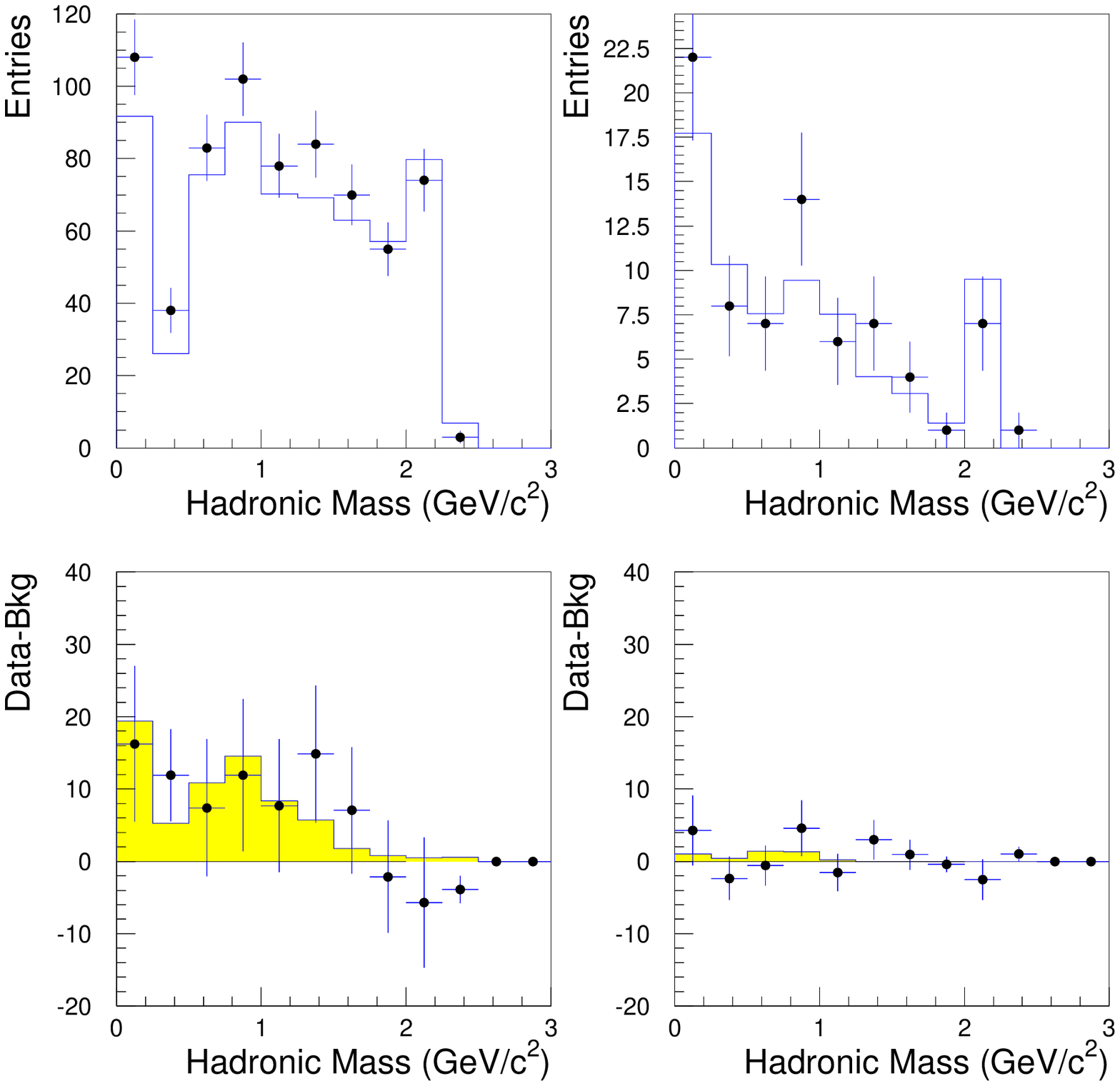,width=14.0cm,height=14.0cm}
\end{center}
\caption[]{\sl Invariant mass $M_X$ distributions for $b \rightarrow u$ 
enriched decays with a secondary hadronic system consisting of either a single 
charged particle or two particles forming a neutral or unit-charge 
secondary system for opposite sign and neutral-charge (left plots) and 
same-sign (right plots) hadronic-lepton system. 
The upper histograms show the expected distribution 
from backgrounds, the points with error bars the data. In the lower
histograms the background subtracted data are compared with the expected
distribution from signal $B \rightarrow X_u \ell \bar \nu$ events.}
\label{fig:8}
\end{figure}

\end{document}